\documentclass[fleqn,usenatbib,useAMS]{mnras}
 \usepackage{graphicx}
 \usepackage{amsmath}
 \usepackage{amssymb}
 \usepackage[T1]{fontenc}
 \usepackage{ae,aecompl}
 \usepackage{txfonts}

 \title[On the visibility of Planet Nine]
       {Finding Planet Nine: a Monte Carlo approach}

 \author[C. de la Fuente Marcos and R. de la Fuente Marcos]
        {C.~de~la~Fuente~Marcos\thanks{E-mail: carlosdlfmarcos@gmail.com}
         and
         R. de la Fuente Marcos \\
         Apartado de Correos 3413, E-28080 Madrid, Spain}
 \date{Accepted 2016 March 21.
       Received 2016 March 21;
       in original form 2016 January 25}
 \pubyear{2016}
 
\begin{document}
  \label{firstpage}
  \pagerange{\pageref{firstpage}--\pageref{lastpage}}
  \maketitle

  \begin{abstract}
     Planet Nine is a hypothetical planet located well beyond Pluto that 
     has been proposed in an attempt to explain the observed clustering in 
     physical space of the perihelia of six extreme trans-Neptunian objects 
     or ETNOs. The predicted approximate values of its orbital elements 
     include a semimajor axis of 700 au, an eccentricity of 0.6, an 
     inclination of 30\degr, and an argument of perihelion of 150\degr. 
     Searching for this putative planet is already under way. Here, we use 
     a Monte Carlo approach to create a synthetic population of Planet Nine 
     orbits and study its visibility statistically in terms of various 
     parameters and focusing on the aphelion configuration. Our analysis 
     shows that, if Planet Nine exists and is at aphelion, it might be 
     found projected against one out of four specific areas in the sky. 
     Each area is linked to a particular value of the longitude of the 
     ascending node and two of them are compatible with an apsidal 
     anti-alignment scenario. In addition and after studying the current 
     statistics of ETNOs, a cautionary note on the robustness of the 
     perihelia clustering is presented. 
  \end{abstract}


  \section{Introduction}
     Batygin \& Brown (2016) have predicted the existence of a massive planet well beyond Pluto in order to explain the observed clustering 
     in physical space of the perihelia of the extreme trans-Neptunian objects or ETNOs (90377) Sedna, 2004~VN$_{112}$, 2007~TG$_{422}$, 
     2010~GB$_{174}$, 2012~VP$_{113}$, and 2013~RF$_{98}$. Such clustering is fairly obvious in terms of the values of their arguments of 
     perihelion in Table \ref{etnos} and, even more clear, positionally in Table \ref{discovery} and Fig. \ref{hunt2} at about 3$^{\rm h}$ 
     in right ascension, $\alpha$, and 0\degr in declination, $\delta$. The putative object responsible for inducing this clustering has 
     been provisionally denominated Planet Nine. Batygin \& Brown (2016) have provided tentative values for the orbital parameters of the 
     proposed 10 Earth masses planet (semimajor axis, $a=700$~au, eccentricity, $e=0.6$, inclination, $i=30$\degr, and argument of 
     perihelion, $\omega=150$\degr) and discussed its possible location in the sky.\footnote{http://www.findplanetnine.com} Planet Nine as 
     characterized by Batygin \& Brown (2016) is expected to exhibit a magnitude $V$ in the range 16--21 at perihelion and 20--25 at 
     aphelion. Detailed modelling by Linder \& Mordasini (2016) gives a magnitude $V$ from the reflected light of 23.7 at aphelion. 

     In principle, searching for this putative planet is feasible for currently active moving object surveys and it is already under way. 
     For slow-moving objects, most active surveys can record candidates brighter than $V=22$~mag (see e.g. Harris \& D'Abramo 2015). The 
     faintest natural moving object observation performed so far corresponds to $V=26.7$~mag for asteroid 2008 LG$_{2}$ (Micheli et al. 
     2015) and the faintest objects observable with 8-m class telescopes have about 27. V774104 was discovered with magnitude 24 at 103 au 
     (Sheppard, Trujillo \& Tholen 2015). Detection of moving objects not only depends on their apparent magnitude but on their rate of 
     motion as well (see e.g. Harris \& D'Abramo 2015). However, with an average daily motion of nearly 1 arcsec d$^{-1}$ at perihelion and 
     almost 0.06 arcsec d$^{-1}$ at aphelion, this should be a non-issue for Planet Nine; the object's sky motion is far too slow. 
     Fortunately, its shifting with respect to background stars due to parallax as the Earth moves around the Sun could be as high as 
     3 arcsec d$^{-1}$ at aphelion (V774104 was identified via parallax, not daily motion). In any case, if Planet Nine currently moves 
     projected against a rich stellar background (a bright section of the Milky Way galaxy for instance) and/or its apparent magnitude is 
     $>22$ in $V$, its eventual identification could be particularly challenging. 

     The Planet Nine hypothesis presents a suitable and robust scenario to explain the orbital properties of six ETNOs, but it may not be 
     adequate to account for the apparent clustering of arguments of perihelion around 0\degr (Trujillo \& Sheppard 2014) and inclination 
     around 20\degr (de la Fuente Marcos \& de la Fuente Marcos 2014) observed for ETNOs with values of the semimajor axis in the range 
     150--250~au. A number of scenarios aimed at explaining the available observational evidence have been proposed since the discovery of 
     2012~VP$_{113}$ (Trujillo \& Sheppard 2014). They include the possible existence of one (Trujillo \& Sheppard 2014; Gomes, Soares \& 
     Brasser 2015; Malhotra, Volk \& Wang 2016) or more trans-Plutonian planets (de la Fuente Marcos \& de la Fuente Marcos 2014; de la 
     Fuente Marcos, de la Fuente Marcos \& Aarseth 2015), capture of ETNOs within the Sun's natal open star cluster (J\'{\i}lkov\'a et al. 
     2015), stellar encounters (Brasser \& Schwamb 2015; Feng \& Bailer-Jones 2015), being a by-product of Neptune's migration (Brown \& 
     Firth 2016) or the inclination instability (Madigan \& McCourt 2016), and being the result of Milgromian dynamics (Pau\v{c}o \& 
     Kla\v{c}ka 2016). In any case, trans-Plutonian planets ---if they do exist--- cannot be too massive or bright (Iorio 2014; Luhman 2014; 
     Cowan, Holder \& Kaib 2016; Fienga et al. 2016; Ginzburg, Sari \& Loeb 2016; Linder \& Mordasini 2016) to have escaped detection during 
     the last two decades of surveys and astrometric studies; masses close to or below those of Uranus or Neptune are most likely. 
     Trans-Plutonian planets may have been scattered out of the region of the giant planets early in the history of the Solar system (see 
     e.g. Bromley \& Kenyon 2014) or even captured from another planetary system (Li \& Adams 2016), but planets similar to Uranus or 
     Neptune (super-Earths) may also form at 125--250 au from the Sun (Kenyon \& Bromley 2015). The putative existence of trans-Plutonian 
     planets may have a role on models aimed at explaining periodic mass extinctions (Whitmire 2016).

     The study of the visibility of the ETNOs carried out in de la Fuente Marcos \& de la Fuente Marcos (2014) revealed an intrinsic bias in 
     declination induced by our observing point on Earth: the vast majority must reach perihelion (i.e. perigee) at declinations in the 
     range $-$24\degr to 24\degr. Here, we study the visibility of a synthetic population of Planet Nine virtual orbits from the Earth to 
     uncover possible biases that may affect the detectability of such object if it exists. This Letter is organized as follows. Section~2 
     is a review of ETNOs statistics that includes a cautionary note regarding the perihelia clustering identified in Batygin \& Brown 
     (2016). Our Monte Carlo methodology is briefly reviewed in Section~3. The distribution in equatorial coordinates of Planet Nine virtual 
     orbits at aphelion is studied in Section~4. Section~5 repeats the analysis for the location of Planet Nine in Fienga et al. (2016). 
     Results are discussed in Section~6 and conclusions are summarized in Section~7. 
%
%
      \begin{table*}
        \centering
        \fontsize{8}{11pt}\selectfont
        \tabcolsep 0.10truecm
        \caption{Various orbital parameters ---$q=a(1-e)$, $Q=a(1+e)$, $\varpi=\Omega+\omega$, $P$ is the orbital period, $\Omega^*$ and 
                 $\omega^*$ are $\Omega$ and $\omega$ in the interval ($-\pi$, $\pi$) instead of the regular (0, 2$\pi$)--- for the 16 
                 objects discussed in this Letter. The statistical parameters are Q$_{1}$, first quartile, Q$_{3}$, third quartile, IQR, 
                 interquartile range, OL, lower outlier limit (Q$_{1}-1.5$IQR), and OU, upper outlier limit (Q$_{3}+1.5$IQR); see the text 
                 for additional details. (Epoch: 2457400.5, 2016-January-13.0 00:00:00.0 UT. J2000.0 ecliptic and equinox. Source: Jet 
                 Propulsion Laboratory Small-Body Database. Data retrieved on 27 February 2016.)
                }
        \begin{tabular}{lrrrrrrrrrrr}
          \hline
             Object             & $a$ (au) & $e$      & $i$ (\degr) & $\Omega$ (\degr) & $\omega$ (\degr) & $\varpi$ (\degr) & $q$ (au) & 
                       $Q$ (au) & $P$ (yr)  & $\Omega^*$ (\degr) & $\omega^*$ (\degr) \\
          \hline
        (82158) 2001 FP$_{185}$ & 226.3448 & 0.8486685 & 30.75720   & 179.3004         &   6.9787         & 186.2791         & 34.2531  &
                       418.4364 &  3405.367 &  179.3004          &    6.9787          \\
        (90377) Sedna           & 507.5603 & 0.8501824 & 11.92872   & 144.5463         & 311.4614         &  96.0077         & 76.0415  &
                       939.0792 & 11435.094 &  144.5463          &  $-$48.5386        \\
       (148209) 2000 CR$_{105}$ & 227.9513 & 0.8057223 & 22.71773   & 128.2463         & 317.2193         &  85.4656         & 44.2859  &
                       411.6168 &  3441.687 &  128.2463          &  $-$42.7807        \\
       (445473) 2010 VZ$_{98}$  & 152.7794 & 0.7753635 &  4.50950   & 117.4524         & 313.8953         &  71.3477         & 34.3198  &
                       271.2389 &  1888.449 &  117.4524          &  $-$46.1047        \\
                2002 GB$_{32}$  & 215.7621 & 0.8362043 & 14.17368   & 176.9791         &  36.9855         & 213.9646         & 35.3409  &
                       396.1834 &  3169.356 &  176.9791          &   36.9855          \\
                2003 HB$_{57}$  & 164.6181 & 0.7685925 & 15.47644   & 197.8293         &  10.7805         & 208.6098         & 38.0939  &
                       291.1424 &  2112.149 & -162.1707          &   10.7805          \\
                2003 SS$_{422}$ & 193.8328 & 0.7966122 & 16.80783   & 151.1119         & 209.8843         &   0.9962         & 39.4232  &
                       348.2424 &  2698.666 &  151.1119          & $-$150.1157        \\
                2004 VN$_{112}$ & 321.0199 & 0.8525664 & 25.56295   &  66.0107         & 327.1707         &  33.1814         & 47.3291  &
                       594.7106 &  5751.830 &   66.0107          &  $-$32.8293        \\
                2005 RH$_{52}$  & 151.1376 & 0.7420410 & 20.46234   & 306.1711         &  32.3890         & 338.5601         & 38.9873  &
                       263.2879 &  1858.091 &  -53.8289          &   32.3890          \\
                2007 TG$_{422}$ & 492.7277 & 0.9277916 & 18.58697   & 112.9515         & 285.7968         &  38.7483         & 35.5791  &
                       949.8764 & 10937.517 &  112.9515          &  $-$74.2032        \\
                2007 VJ$_{305}$ & 188.3373 & 0.8131705 & 12.00306   &  24.3834         & 338.3611         &   2.7445         & 35.1870  &
                       341.4876 &  2584.715 &   24.3834          &  $-$21.6389        \\
                2010 GB$_{174}$ & 371.1183 & 0.8687090 & 21.53812   & 130.6119         & 347.8124         & 118.4243         & 48.7245  &
                       693.5121 &  7149.518 &  130.6119          &  $-$12.1876        \\
                2012 VP$_{113}$ & 259.3002 & 0.6896024 & 24.04680   &  90.8179         & 293.7168         &  24.5346         & 80.4862  &
                       438.1142 &  4175.538 &   90.8179          &  $-$66.2832        \\
                2013 GP$_{136}$ & 152.4968 & 0.7303547 & 33.48578   & 210.7142         &  42.1284         & 252.8426         & 41.1201  &
                       263.8736 &  1883.213 & -149.2858          &   42.1284          \\
                2013 RF$_{98}$  & 309.0738 & 0.8826022 & 29.61402   &  67.5205         & 316.4991         &  24.0196         & 36.2846  &
                       581.8631 &  5433.774 &   67.5205          &  $-$43.5009        \\
                2015 SO$_{20}$  & 162.7035 & 0.7961710 & 23.44153   &  33.6221         & 354.9699         &  28.5920         & 33.1637  &
                       292.2434 &  2075.409 &   33.6221          &   $-$5.0301        \\
          \hline
             Mean               & 256.0478 & 0.8115221 & 20.31954   & 133.6418         & 221.6281         & 107.7699         & 43.6637  &
                       468.4318 &  4375.023 &   66.1418          &  $-$25.8719        \\ 
             Standard deviation & 115.6941 & 0.0616087 &  7.71647   &  71.9552         & 140.3086         & 102.2955         & 14.3016  &
                       225.1645 &  3077.676 &  105.7150          &   48.9934          \\
             Median             & 221.0535 & 0.8094464 & 21.00023   & 129.4291         & 302.5891         &  78.4066         & 38.5406  &
                       403.9001 &  3287.362 &  101.8847          &  $-$27.2341        \\
             Q$_{1}$            & 164.1395 & 0.7736707 & 15.15075 &  84.9935           &  40.8427         &  27.5777         & 35.3024  &
                       291.9681 &  2102.964 &   31.3124          &  $-$46.7132        \\
             Q$_{3}$            & 312.0603 & 0.8507784 & 24.42584 & 177.5594           & 319.7071         & 191.8618         & 45.0467  &
                       585.0750 &  5513.288 &  134.0955          &    7.9291          \\
             IQR                & 147.9209 & 0.0771077 &  9.27509 &  92.5659           & 278.8645         & 164.2841         &  9.7443  &
                       293.1068 &  3410.324 &  102.7831          &   54.6423          \\
             OL                 & $-$57.7418 & 0.6580092 &  1.23812 & $-$53.8553         &$-$377.4541       &$-$218.8485       & 20.6860  &
                    $-$147.6921 & $-$3012.521 & $-$122.8622          & $-$128.6766        \\
             OU                 & 533.9416 & 0.9664399 & 38.33846 & 316.4082           & 738.0039         & 438.2880         & 59.6631  &
                      1024.7352 & 10628.773 &  288.2701          &   89.8926          \\
          \hline
        \end{tabular}
        \label{etnos}
      \end{table*}
%
%
%
%
      \begin{table}
        \centering
        \fontsize{8}{11pt}\selectfont
        \tabcolsep 0.05truecm
        \caption{Equatorial coordinates, apparent magnitudes (with filter if known) at discovery time, absolute magnitude, and $\omega$ for
                 the 16 objects discussed in this Letter. (J2000.0 ecliptic and equinox. Source: Minor Planet Center ---MPC--- Database.)
                }
        \begin{tabular}{lcclcr}
          \hline
             Object          & $\alpha$ ($^{\rm h}$:$^{\rm m}$:$^{\rm s}$) & $\delta$ (\degr:\arcmin:\arcsec) & $m$ (mag)  & $H$ (mag) & $\omega$ (\degr) \\
          \hline
             82158           & 11:57:50.69                                 & +00:21:42.7                      & 22.2  (R)  &  6.0      &   6.77           \\
             Sedna           & 03:15:10.09                                 & +05:38:16.5                      & 20.8  (R)  &  1.5      & 311.19           \\
             148209          & 09:14:02.39                                 & +19:05:58.7                      & 22.5  (R)  &  6.3      & 317.09           \\
             445473          & 02:08:43.575                                & +08:06:50.90                     & 20.3  (R)  &  5.0      & 313.80           \\
             2002~GB$_{32}$  & 12:28:25.94                                 & $-$00:17:28.4                    & 21.9  (R)  &  7.7      &  36.89           \\
             2003~HB$_{57}$  & 13:00:30.58                                 & $-$06:43:05.4                    & 23.1  (R)  &  7.4      &  10.64           \\
             2003~SS$_{422}$ & 23:27:48.15                                 & $-$09:28:43.4                    & 22.9  (R)  &  7.1      & 209.98           \\
             2004~VN$_{112}$ & 02:08:41.12                                 & $-$04:33:02.1                    & 22.7  (R)  &  6.4      & 327.23           \\
             2005~RH$_{52}$  & 22:31:51.90                                 & +04:08:06.1                      & 23.8  (g)  &  7.8      &  32.59           \\
             2007~TG$_{422}$ & 03:11:29.90                                 & $-$00:40:26.9                    & 22.2       &  6.2      & 285.84           \\
             2007~VJ$_{305}$ & 00:29:31.74                                 & $-$00:45:45.0                    & 22.4       &  6.6      & 338.53           \\
             2010~GB$_{174}$ & 12:38:29.365                                & +15:02:45.54                     & 25.09 (g)  &  6.5      & 347.53           \\
             2012~VP$_{113}$ & 03:23:47.159                                & +01:12:01.65                     & 23.1  (r)  &  4.1      & 293.97           \\
             2013~GP$_{136}$ & 14:09:40.000                                & $-$11:30:08.47                   & 23.5  (r)  &  6.6      &  42.16           \\
             2013~RF$_{98}$  & 02:29:07.61                                 & $-$04:56:34.6                    & 23.5  (z)  &  8.6      & 316.55           \\
             2015~SO$_{20}$  & 01:01:17.301                                & $-$03:11:00.81                   & 21.4  (R)  &  6.4      & 354.97           \\
          \hline
        \end{tabular}
        \label{discovery}
      \end{table}
%
%

  \section{ETNOs: current statistics}
     In Trujillo \& Sheppard (2014) the ETNOs are defined as asteroids with semimajor axis greater than 150 au and perihelion greater than 
     30 au. At present, there are 16 known ETNOs (see Tables \ref{etnos} and \ref{discovery} for relevant data). The descriptive statistics 
     of this sample are included in Table \ref{etnos}; in this table, unphysical values are displayed for completeness. From these results 
     it is obvious that the strongest clustering is observed in $e$ and $i$. As pointed out in de la Fuente Marcos \& de la Fuente Marcos 
     (2014), the clustering in $e$ can be the result of observational bias but the one in $i$ cannot be explained as resulting from 
     selection effects, it must have a dynamical origin. As the one in $e$, the clustering in the values of the perihelion distance may be
     explained as a selection effect. It is also clear that the new additions to the ETNO group since the discovery of 2012 VP$_{113}$ 
     follow the trends already identified in 2014; in particular, the orbits of 2004~VN$_{112}$ and 2013~RF$_{98}$ are alike. However, we 
     would like to point out a potentially important issue even if the ETNO sample is still small. In statistics, outliers are often defined 
     as observations that fall below Q$_{1}-1.5$ IQR or above Q$_{3}+1.5$ IQR, where Q$_{1}$ is the first or lower quartile, Q$_{3}$ is the 
     third or upper quartile, and IQR is the interquartile range or difference between the upper and lower quartiles. In general, there are 
     no outliers among the ETNOs (e.g. 2003 SS$_{422}$ is an outlier in terms of $\omega^*$, see Table \ref{etnos}), but both (90377) Sedna 
     and 2012 VP$_{113}$ are statistical outliers in terms of perihelion distance, $q$. The upper boundary for outliers in $q$ is 59.7~au; 
     the values of the perihelion distance of both Sedna and 2012 VP$_{113}$ are well above this upper limit. Sedna is also an outlier in 
     terms of orbital period. The sample is small but this fact may be signalling a different dynamical context for these two objects. It is 
     statistically possible that Sedna and 2012 VP$_{113}$ are members of a separate dynamical class within the ETNOs and therefore be 
     subjected to a different set of perturbations. 

     The scenario in which trans-Plutonian planets keep the values of the orbital parameters of the ETNOs in check thanks to a particular 
     case of the Kozai mechanism (Kozai 1962) discussed in Trujillo \& Sheppard (2014), de la Fuente Marcos \& de la Fuente Marcos (2014) or 
     de la Fuente Marcos et al. (2015) and the resonant coupling mechanism described in Batygin \& Brown (2016) create dynamical pathways
     that in some cases may deliver objects to high inclination or even retrograde orbits. In principle, the mechanism detailed in de la 
     Fuente Marcos et al. (2015) can also produce objects with orbits at steeply inclined angles, it already does it for Jupiter. On the 
     other hand, the scenarios described in de la Fuente Marcos et al. (2015) and Batygin \& Brown (2016) are not incompatible, and the 
     Kozai mechanism can operate at high eccentricities (see e.g. Naoz 2016) when the value of the relative longitude of perihelion, 
     $\Delta\varpi$, librates about 180\degr (apsidal anti-alignment). A hypothetical Planet Nine may induce Kozai-like behaviour, i.e. 
     libration of the value of the argument of perihelion of the ETNOs that reproduces the observed clustering. The currently known ETNOs 
     probably represent evolutionary steps within dynamical tracks and some of them are more dynamically evolved than others. The presence 
     of statistical outliers could be a sign of this.

  \section{A Monte Carlo approach}
     The study of the visibility of a set of orbits with parameters defined within some boundary values is a statistical problem well suited
     to apply Monte Carlo techniques (Metropolis \& Ulam 1949; Press et al. 2007). A representative sample of the set of orbits under study
     is systematically explored so the regions in the sky with optimal visibility (highest probability) can be determined; in our case, the 
     equations of the orbits of both the Earth and Planet Nine under the two-body approximation (e.g. Murray \& Dermott 1999) are sampled to 
     find the minimum and maximum distance between the Earth and Planet Nine. This technique was used in de la Fuente Marcos \& de la Fuente 
     Marcos (2014) to analyse the visibility of the ETNOs, and the details and further references are given there. Using a Monte Carlo 
     approach, we generate a synthetic population of Planet Nines with semimajor axis, $a\in$ (650, 750) au, eccentricity, $e\in$ (0.55, 
     0.65), inclination, $i\in$ (25, 35)\degr, longitude of the ascending node, $\Omega\in$ (0, 360)\degr, and argument of perihelion, 
     $\omega\in$ (140, 160)\degr as no explicit value of $\Omega$ is given in Batygin \& Brown (2016). We assume that the orbits of the 
     multiple instances of Planet Nine are uniformly distributed in orbital parameter space. Ten million test orbits have been studied 
     focusing on the visibility at aphelion. The analyses in Fienga et al. (2016) and Linder \& Mordasini (2016) strongly disfavour a 
     present-day Planet Nine located at perihelion.
%
%
      \begin{figure}
        \centering
         \includegraphics[width=\linewidth]{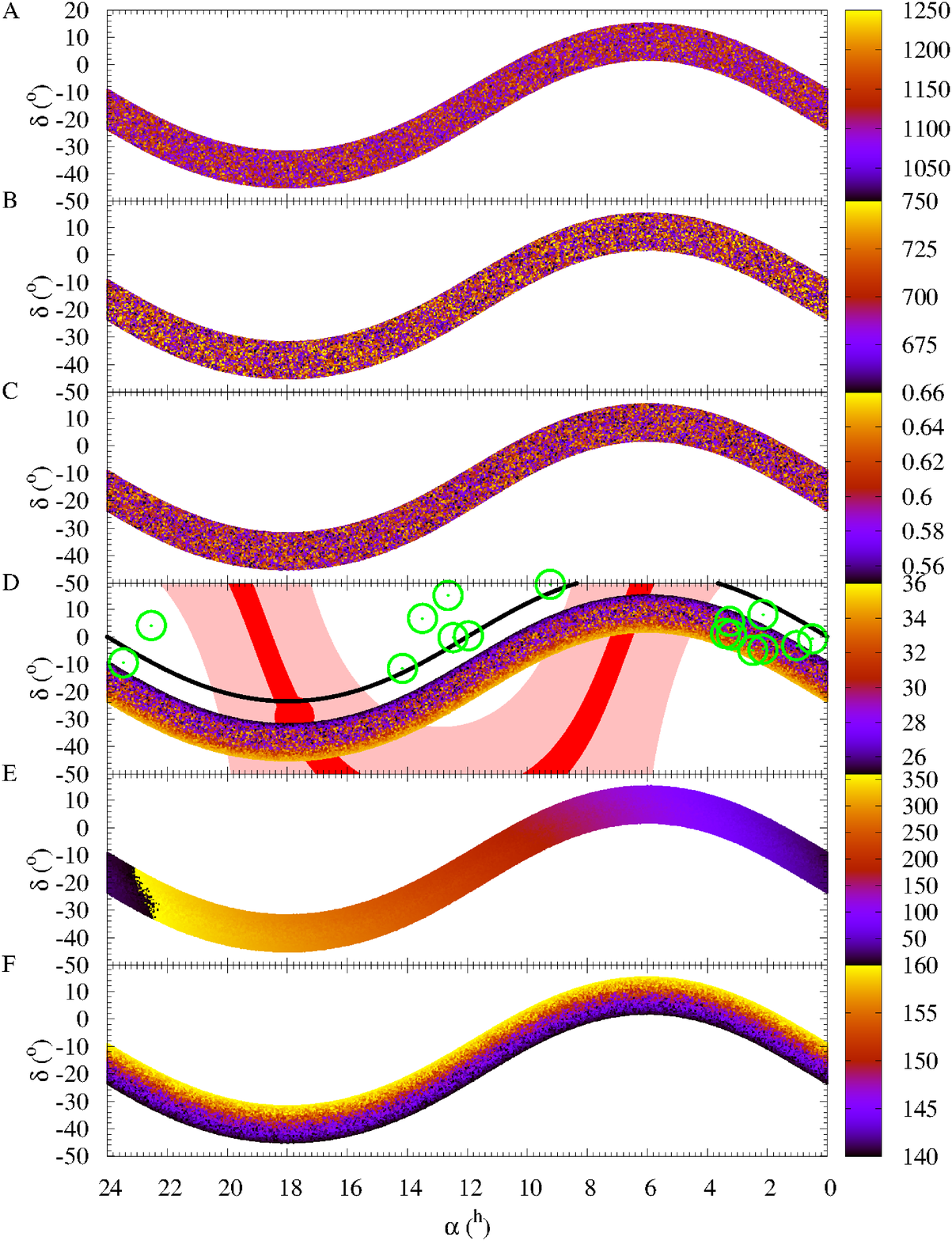}
         \caption{Distribution in equatorial coordinates of the studied orbits at aphelion as a function of various orbital elements and 
                  parameters. As a function of the value of the aphelion distance (panel A), of $a$ (panel B), of $e$ (panel C), of $i$ 
                  (panel D), of $\Omega$ (panel E), and of $\omega$ (panel F). The green circles in panel D give the location at discovery 
                  of all the known ETNOs. Also in panel D, the Galactic disc is arbitrarily defined as the region confined between galactic 
                  latitude $-5$\degr and 5\degr (in red), the position of the Galactic Centre is represented by a filled red circle, the 
                  region enclosed between galactic latitude $-30$\degr and 30\degr appears in pink, and the ecliptic appears in black.
                 }
         \label{hunt2}
      \end{figure}
%
%

     The distribution in equatorial coordinates of the set of studied orbits is presented in Fig. \ref{hunt2}. In this figure, the value of 
     the parameter in the appropriate units is colour coded following the scale printed on the associated colour box. In panel D 
     (inclination), the locations of the Galactic disc and centre are indicated. The background stellar density is the highest towards these 
     regions in the sky. The distribution of aphelion distances, semimajor axes and eccentricities is rather uniform. The distribution in 
     inclination and argument of perihelion depends on the declination; those orbits with higher values of the inclination reach aphelion at 
     lower declinations, the same behaviour is observed for the ones with lower values of the argument of perihelion. The distribution in 
     longitude of the ascending node depends on the right ascension; orbits with $\Omega\sim$0\degr reach aphelion at $\alpha\sim$23$^{\rm 
     h}$, the ones with $\Omega\sim$ 90\degr at $\alpha\sim$5$^{\rm h}$, and those with $\Omega\sim$270\degr reach aphelion at 
     $\alpha\sim$17$^{\rm h}$. In any case, orbits reach perihelion at declination in the range $-$20\degr to 50\degr (not shown) and 
     aphelion in the range $-$50\degr to 20\degr. This is markedly different from the bias found for the ETNOs in de la Fuente Marcos \& de 
     la Fuente Marcos (2014); i.e. searching for Planet Nine is not expected to increase the discovery rate of ETNOs and searching for ETNOs 
     is not going to make a direct detection of Planet Nine more likely (indirectly yes, by improving the values of its putative orbital 
     elements).

  \section{Planet Nine at aphelion}
\begin{figure}
        \centering
         \includegraphics[width=\linewidth]{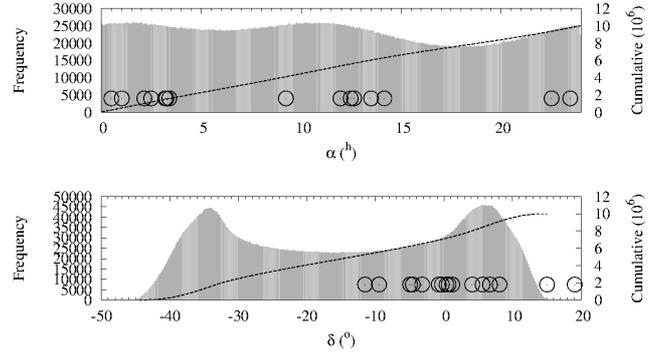}
         \caption{Frequency distribution in equatorial coordinates (right ascension, top panel, and declination, bottom panel) of the
                  studied virtual orbits at aphelion. The distribution is somewhat uniform in right ascension and shows two maxima for 
                  declinations in the ranges ($-$40, $-$30)\degr and (0, 10)\degr. In this and subsequent figures, the number of bins is 
                  2 $n^{1/3}$, where $n$ is the number of virtual orbits plotted, error bars are too small to be seen. The black circles 
                  correspond to objects in Table \ref{discovery}.
                 }
         \label{radeca}
      \end{figure}

\begin{figure*}
        \centering
         \includegraphics[width=0.32\linewidth]{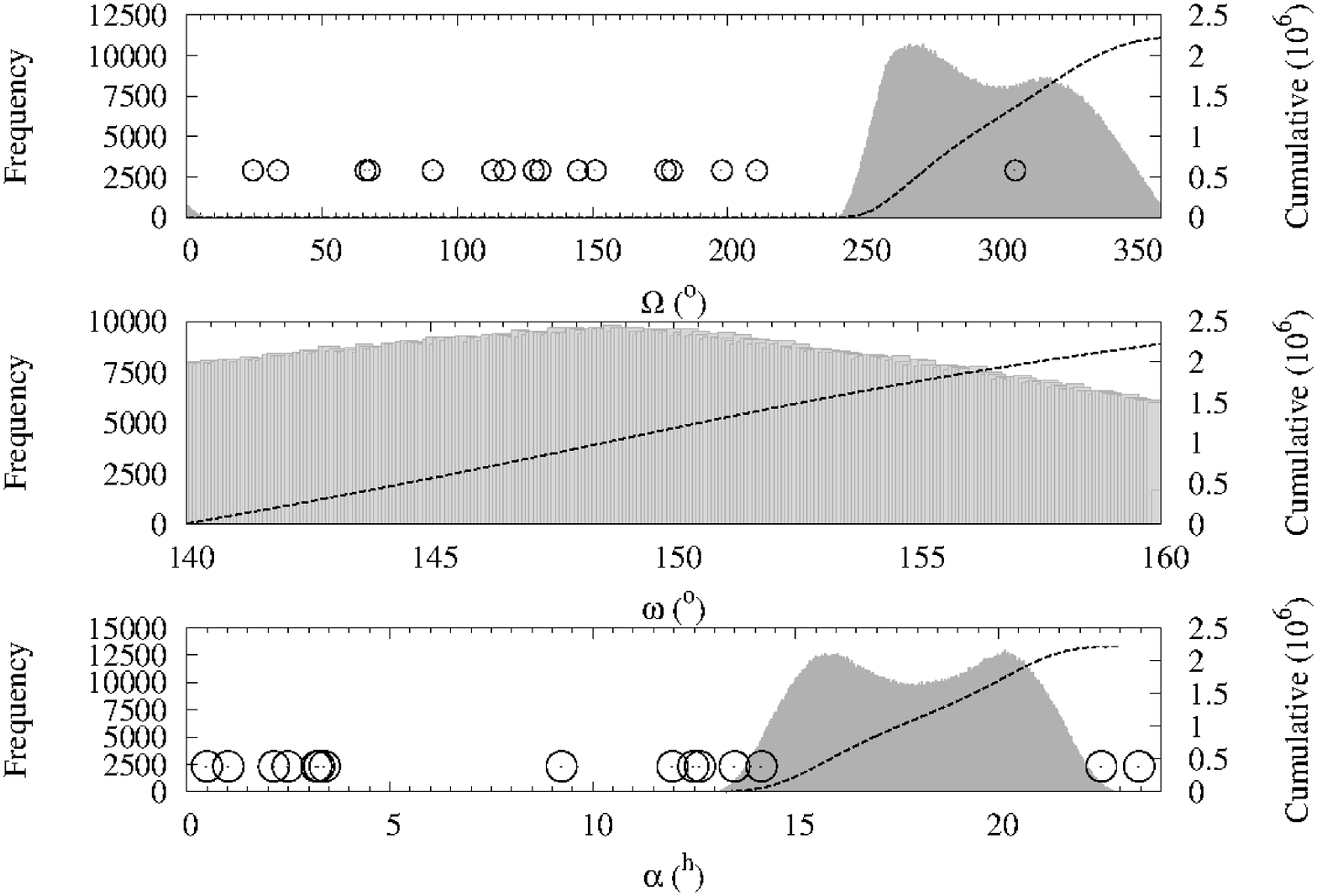}
         \includegraphics[width=0.32\linewidth]{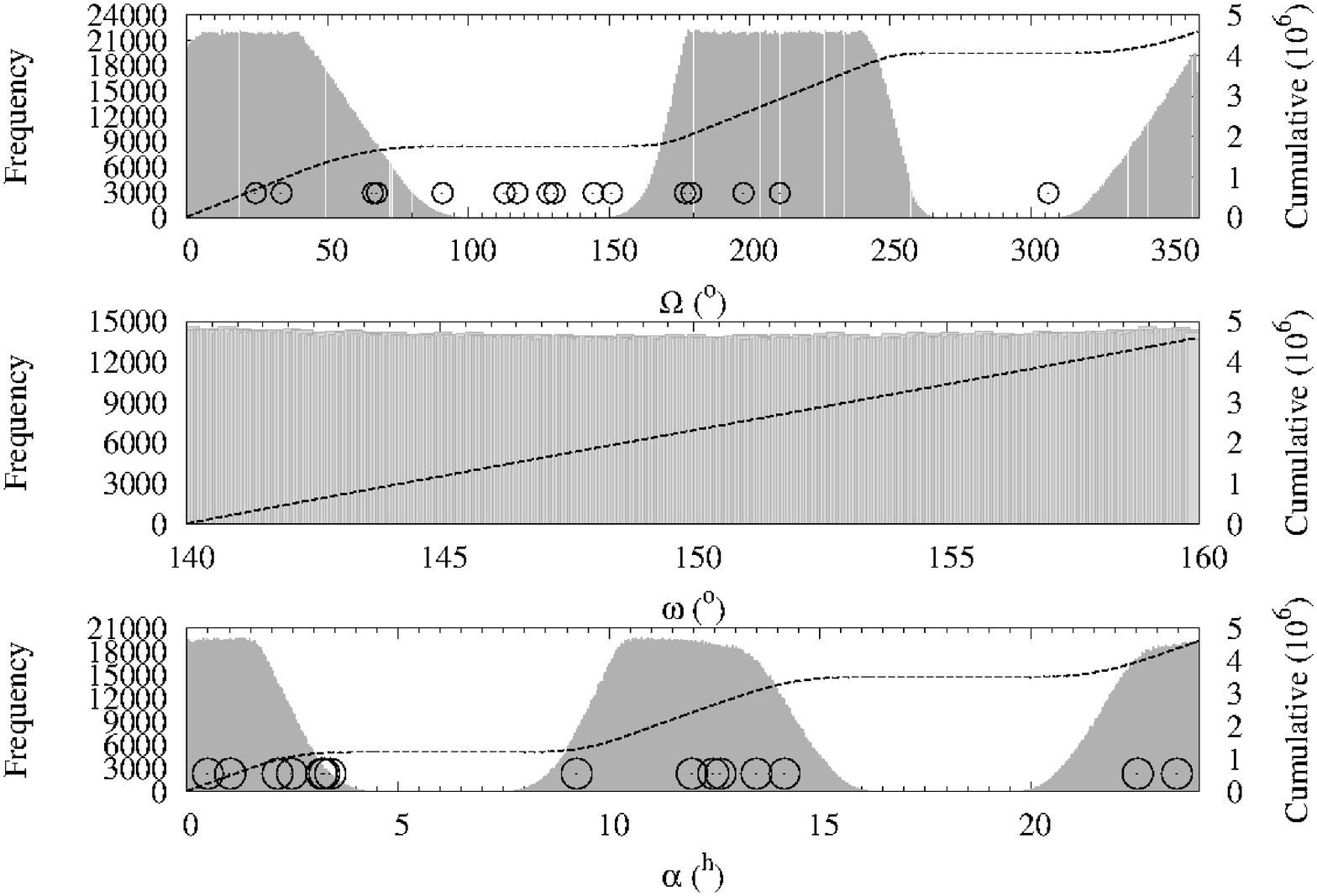}
         \includegraphics[width=0.32\linewidth]{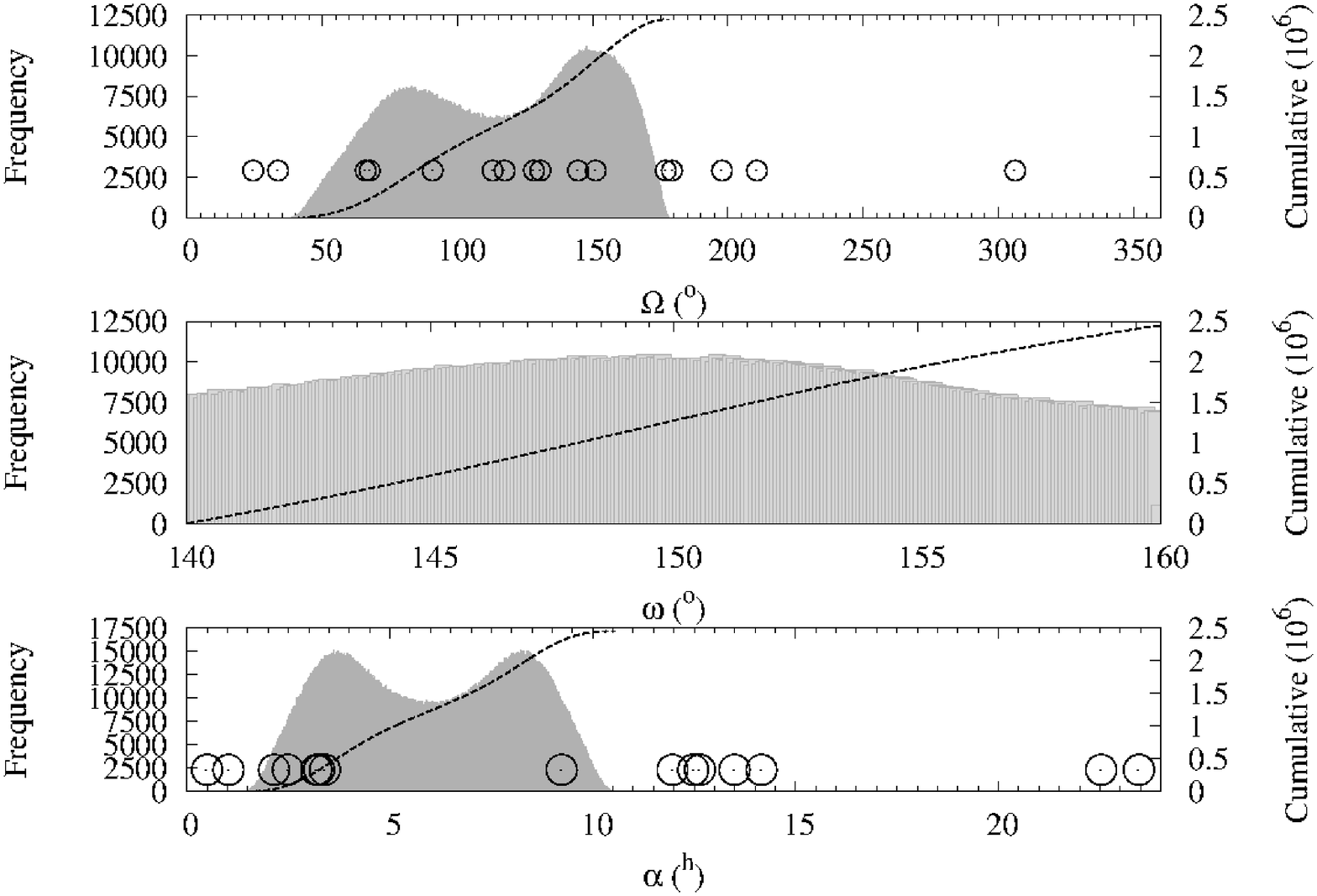}\\
         \caption{Frequency distribution in $\Omega$ (top panels), $\omega$ (middle panels), and right ascension (bottom panels) of virtual 
                  orbits with $\delta$ in the ranges ($-$40, $-$30)\degr (left-hand panels), ($-$30, 0)\degr (central panels), and 
                  (0, 10)\degr (right-hand panels). The black circles correspond to objects in Tables \ref{etnos} and \ref{discovery}. 
                 }
         \label{regionsa}
      \end{figure*}

     Detection of Planet Nine would be much easier if it is close to perihelion at present, but the analysis of Cassini radio ranging data
     carried out in Fienga et al. (2016) strongly suggests that Planet Nine as characterized by Batygin \& Brown (2016) cannot be at 
     perihelion. This conclusion is consistent with the analysis of its expected photometric properties in e.g. Linder \& Mordasini (2016). 
     A location in or near aphelion is however compatible with both the analysis of planetary ephemerides and the outcome of the many 
     surveys completed in recent years. If Planet Nine is currently near aphelion, its declination will be in the range $-$50\degr to 
     20\degr (see Fig. \ref{hunt2}), but not all the values of $\delta$ are equally probable. Figure \ref{radeca} shows that the 
     distribution is somewhat uniform in right ascension, but the probability of finding an orbit reaching aphelion at declinations in the 
     ranges ($-$40, $-$30)\degr and (0, 10)\degr is nearly 1.7 times higher than that of doing it in the range ($-$30, 0)\degr. The effect 
     of the bias in declination is analysed in more detail in Fig. \ref{regionsa}. Locations near the Galactic Centre are possible if 
     $\Omega\sim300\degr$; the Galactic Centre is approximately at $\alpha\sim17\fh7$ and $\delta\sim-29\degr$. A more exhaustive analysis 
     shows that if $\delta\in(5, 7)\degr$ then $\Omega\in(145, 155)\degr$ and $\alpha\in(8, 9)^{\rm h}$, which is located in Hydra and away 
     from the main bulk of the Milky Way. Unfortunately, another optimal solution is possible ---albeit slightly less probable--- if 
     $\delta\in(-35, -33)\degr$ then $\Omega\in(265, 275)\degr$ and $\alpha\in(15.5, 16.5)^{\rm h}$, which is located towards the Galactic 
     bulge between the constellations of Scorpius and Lupus, but relatively far from the Galactic Centre. These two areas are associated 
     with values of $\Omega$ that give $\Delta\varpi$ equal to 180\degr ($\Omega=150$\degr) and 60\degr ($\Omega=270$\degr), respectively. 
     This second value is inconsistent with the pseudo resonant scenario described in Batygin \& Brown (2016). Figures \ref{radeca} and 
     \ref{regionsa} show that two other solutions are possible but less probable. One with $\delta\in(5, 7)\degr$, $\Omega\sim100$\degr and 
     $\alpha\in(3.5, 4.5)^{\rm h}$ ---in Taurus--- and the second one with $\delta\in($-$35, $-$33)\degr$, $\Omega\sim310$\degr and 
     $\alpha\in(20, 21)^{\rm h}$ ---between Microscopium and Sagittarius. The solution with $\Omega\sim100$\degr is favoured by Batygin and 
     Brown (2016; see footnote 1) and subsequently by Fienga et al. (2016). Such solution gives a value of $\Delta\varpi\sim200$\degr which 
     is also consistent with apsidal anti-alignment. 
%
%
%
%
%
%
%
%

  \section{Cassini hints: Fienga et al. (2016)}
\begin{figure}
        \centering
         \includegraphics[width=\linewidth]{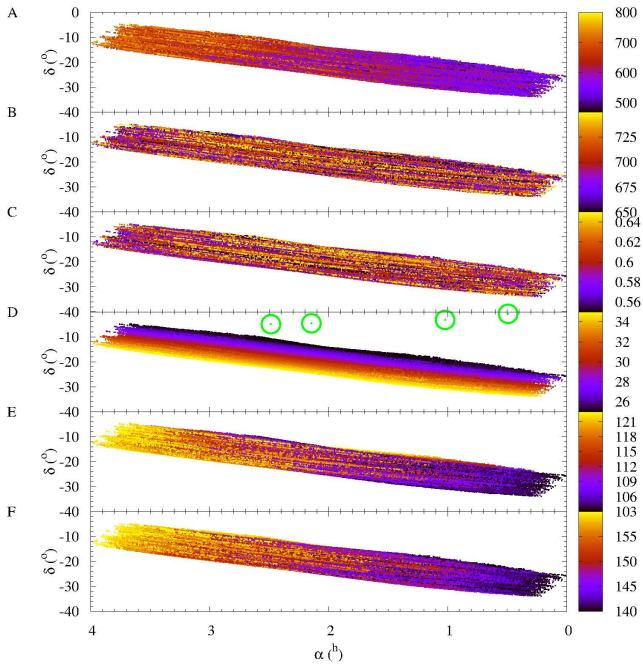}
         \caption{As Fig. \ref{hunt2} but for the orbital solution in Fienga et al. (2016). Here, panel A shows the value of the geocentric 
                  distance. 
                 }
         \label{fiengaetal}
      \end{figure}

     Fienga et al. (2016) have carried out an analysis of Cassini radio ranging data. Their extrapolation of the Cassini data indicates that
     Planet Nine as characterised by Batygin \& Brown (2016) cannot exist in the interval of true anomaly ($-$132, 106.5)\degr. This 
     automatically excludes the perihelion (see their fig. 6). The aphelion is included in the uncertainty zone where the Cassini data do 
     not provide any constraints, i.e. the residuals are compatible with zero. Fienga et al. (2016) indicate that from the point of view of 
     the Cassini residuals, the most probable position of Planet Nine ---assuming a value of $\Omega=113$\degr--- is at true anomaly equal 
     to 117.8\degr$^{+11\degr}_{-10\degr}$. An analysis similar to that in Section 3 gives Fig. \ref{fiengaetal}; this prediction places 
     Planet Nine at $\alpha\sim2^{\rm h}$ and $\delta\sim-20\degr$, in Cetus.  
%
%
%
%

  \section{Discussion}
     The Planet Nine hypothesis represents an exciting opportunity to survey the outskirts of the Solar system with a purpose and improve 
     our current knowledge of that distant region as well as of testing the correctness of models and long-standing assumptions. Our 
     visibility analyses provide clues on the most probable location of the putative planet given a set of assumed orbital parameters based 
     on the published data. The aphelion configuration gives two preferred present locations with nearly the same degree of probability and 
     two others with lower probability. If any of the assumed parameters is grossly in error, the preferred locations computed via Monte 
     Carlo would be different but perhaps not too far from the values discussed here. Less probable locations are mostly close to the 
     regions where ETNOs have already been found and the lack of detections there suggests that their lower values of the probability are 
     confirmed by the available observational data. Figure \ref{hunt2} shows that for finding ETNOs, the region enclosed between galactic 
     latitude $-30$\degr and 30\degr has been so far carefully avoided. In general, this cannot be the case for any serious search for 
     Planet Nine. 

  \section{Conclusions}
     In this Letter, we have explored the visibility of Planet Nine. This study has been performed using Monte Carlo techniques. In 
     addition, the descriptive statistics of the sample of known ETNOs has been re-examined. Our conclusions can be summarized as follows.
     \begin{itemize}
        \item Observing from the Earth, Planet Nine would reach perihelion at declination in the range $-$20\degr to 50\degr and aphelion 
              in the range $-$50\degr to 20\degr.
        \item If Planet Nine is at aphelion, it is most likely moving within $\alpha\in(8, 9)^{\rm h}$ and $\delta\in(5, 7)\degr$. Another
              solution, $\alpha\in(3.5, 4.5)^{\rm h}$ and $\delta\in(5, 7)\degr$, is less probable. Both locations are compatible with an 
              apsidal anti-alignment scenario.
        \item If Planet Nine is at the location favoured in Fienga et al. (2016), it could be found at $\alpha\sim2^{\rm h}$ and 
              $\delta\sim-20\degr$.
        \item The orbits of known ETNOs exhibit robust clustering in inclination at $i=20\pm8$\degr that cannot be explained by selection
              effects.
        \item Considering the sample of known ETNOs, (90377) Sedna and 2012 VP$_{113}$ are clear statistical outliers in terms of 
              perihelion distance. This may indicate that these two objects form a separate dynamical class within the known ETNOs.
     \end{itemize}

  \section*{Acknowledgements}
     We thank the anonymous referee for his/her constructive and helpful report, G. Carraro and E. Costa for acquiring observations of 
     2015 SO$_{20}$, and S. J. Aarseth, D. P. Whitmire, D. Fabrycky and S. Deen for comments on ETNOs and trans-Plutonian planets. This work 
     was partially supported by the Spanish `Comunidad de Madrid' under grant CAM S2009/ESP-1496. In preparation of this Letter, we made use 
     of the NASA Astrophysics Data System, the ASTRO-PH e-print server and the MPC data server.

  \bsp
  \label{lastpage}
\end{document}